\pdfoutput=1
\documentclass[conference,10pt]{IEEEtran}

\newif\ifanon
\anonfalse

\usepackage[T1]{fontenc}
\usepackage[utf8]{inputenc}
\usepackage[varqu]{zi4}   %
\newif\ifisc

\usepackage{graphicx}
\graphicspath{{figs/}}   %
\usepackage{xcolor}
\usepackage{booktabs}
\usepackage{array}
\usepackage{tabularx}
\usepackage{enumitem}
\usepackage{colortbl}
\newcolumntype{H}[1]{>{\bfseries\raggedright\arraybackslash}p{#1}}
\newcolumntype{Y}{>{\raggedright\arraybackslash}X}
\newcolumntype{W}[1]{>{\hsize=#1\hsize\raggedright\arraybackslash}X}
\definecolor{rowshade}{gray}{0.93}
\newcommand{\shaded}{\rowcolor{rowshade}}
\newcommand{\tablestyle}{%
  \footnotesize
  \setlength{\tabcolsep}{3.5pt}%
  \setlength{\aboverulesep}{0pt}%
  \setlength{\belowrulesep}{0pt}%
  \renewcommand{\arraystretch}{1.35}%
}
\usepackage{listings}
\usepackage{tikz}
\usetikzlibrary{arrows.meta,positioning,fit,backgrounds,calc}
\usepackage{xspace}

\newcommand{\Thinflow}{Thinflow\xspace}

\ifanon
    \newcommand{\Kassel}{ClusterB\xspace}
    \newcommand{\Otus}{ClusterA\xspace}
\else
    \newcommand{\Kassel}{Kassel\xspace}
    \newcommand{\Otus}{Otus\xspace}
\fi

\lstdefinestyle{thinflow}{
  basicstyle=\ttfamily\footnotesize,
  columns=fullflexible,
  keepspaces=true,
  breaklines=false,
  showstringspaces=false,
  frame=single,
  framesep=4pt,
  xleftmargin=4pt,
}
\ifanon
    \newcommand{\AnonOnly}[1]{#1}
\else
    \newcommand{\AnonOnly}[1]{}
\fi

\ifanon
    \newcommand{\RepoFootnote}{\footnote{\label{fn:repo}Repository link withheld for double-blind review.}}
\else
    \newcommand{\RepoFootnote}{\footnote{\label{fn:repo}\url{https://github.com/ums-cpl/task-driven-experimentation}}}
\fi
\newcommand{\RepoFootnoteMark}{\unskip\textsuperscript{\ref*{fn:repo}}}

\isctrue   %
\usepackage[backend=biber,style=ieee,doi=true,url=true,eprint=true,sorting=none]{biblatex}
\AtEveryBibitem{\clearfield{note}}   %
\usepackage{xurl}   %
\usepackage[hidelinks]{hyperref}
\usepackage{pbalance}   %

\makeatletter
\def\@IEEEsectpunct{\ \,}
\makeatother

\newcommand{\PaperTitle}{Thinflow: Organizing HPC Benchmarking Campaigns as Rerunnable Directory Trees}
\newcommand{\PaperKeywords}{HPC benchmarking, scaling studies, provenance}

\begin{document}

\title{\PaperTitle}

\ifanon
\else
    \author{
        \IEEEauthorblockN{Richard Schulze, Ari Rasch}
        \IEEEauthorblockA{%
            \textit{University of M\"unster} \\
            M\"unster, Germany \\
            \{r.schulze $|$ a.rasch\}@uni-muenster.de
        }
        \and
        \IEEEauthorblockN{Anna-Lena Roth, Jonas Posner}
        \IEEEauthorblockA{%
            \textit{Fulda University of Applied Sciences} \\
            Fulda, Germany \\
            \{anna-lena.roth $|$ jonas.posner\}@cs.hs-fulda.de
        }
    }
    \hypersetup{
        pdftitle={\PaperTitle},
        pdfauthor={Richard Schulze, Ari Rasch, Anna-Lena Roth, Jonas Posner},
        pdfkeywords={\PaperKeywords}
    }
\fi

\maketitle

\begin{abstract}
    A benchmarking campaign provides the measurements behind an HPC performance study: code variants are built, swept over core or node counts, and measured repeatedly.
Moving such a campaign to another supercomputer or rerunning it months later requires work that is not part of the study itself.
Job scripts are adapted to each supercomputer, and what exactly ran is often reconstructed afterwards.

We present Thinflow, a small Bash engine for organizing HPC benchmarking campaigns as rerunnable directory trees of tasks and runs.
A task is a directory with a short shell script next to the code.
For each sweep, one command builds the code, runs the benchmarks, and plots the results.
Thinflow derives the execution order from declared dependencies and writes it to a plain-text manifest.
Per-supercomputer profiles submit the runs to a workload manager such as Slurm.
Each run writes a provenance record of its environment, hardware, and parameters.
The engine ships inside the campaign directory, so a campaign stays rerunnable without installing a separate tool and, once completed, may serve as a paper artifact.

In a case study, we ran one campaign with identical task files on three Slurm partitions of two supercomputers, on up to 16 nodes.
This strong-scaling study of matrix multiplication in C/OpenMP, C/MPI, and Julia/MPI comprises 37 task files with 217 non-comment lines.
Its 455 benchmark runs recorded what differed per partition, chiefly profiles and command-line parameters.
This suggests that such a campaign needs little structure beyond plain shell files.

\end{abstract}


\section{Introduction}
\label{sec:intro}
Performance studies in HPC rest on benchmarking campaigns: several variants of a code are built, each variant is swept over core counts, node counts, or problem sizes, every point is repeated to quantify variability, and the measurements end up in the figures of a paper.
Moving such a campaign to another supercomputer, rerunning it months later, or handing it to a coauthor, reviewer, or reader requires work that is not part of the study itself.
Job scripts are adapted to each supercomputer, failed runs are found and resubmitted, and what exactly ran is often reconstructed afterwards.

Sound performance reporting has been a concern for decades, from Bailey's catalog of misleading practices~\cite{bailey1991twelve} and its updates for multicore processors and deep-learning workloads~\cite{hager2010fooling,hoefler2018dltwelve} to the guidelines of Hoefler and Belli~\cite{hoefler2015benchmarking}.
Major HPC venues now review research artifacts and award badges for them~\cite{acmbadging,sc26adae}.
Yet a large share of papers ships no artifact~\cite{frachtenberg2022artifacts}, and a survey of almost 300 papers from 2023 finds that current practice does not address artifact longevity~\cite{guilloteau2024longevity}.

Benchmarking frameworks take over much of the work around a campaign.
JUBE, ReFrame, and Benchpark with Ramble describe experiments in a structured form, keep supercomputer settings apart from benchmark definitions, and automate submission and result collection~\cite{steinbach2026jube,karakasis2020reframe,pearce2023benchpark,jacobsen2023ramble}.
They differ in scope: JUBE targets systematic benchmark studies of any size, ReFrame targets regression and performance testing across supercomputers, and Benchpark shares benchmark and system specifications across sites.
We explore one small corner of the same design space: a single, paper-sized campaign that its authors write, run on a few supercomputers, and publish together with the engine that runs it.
We ask whether a few conventions on top of plain shell files suffice to move such a campaign between supercomputers and to record what ran.

In this paper, we present \Thinflow\RepoFootnote{}, a small Bash engine that organizes such a campaign as a directory tree of \emph{tasks} and \emph{runs}.
A task is a unit of work, such as compiling a code variant, running a benchmark, or plotting results.
A run is one execution of a task, for example one repetition of one point of a parameter sweep.
\Thinflow has the following five properties:
\begin{enumerate}[label=\textbf{P\arabic*}, ref=P\arabic*, leftmargin=*]
    \item\label{p:tree} A task is a directory with a short shell script and optional \texttt{KEY=VALUE} settings, settings are inherited along the tree, and each run has its own output directory.
    \Thinflow derives the execution order from declared dependencies and writes it to a plain-text manifest before anything is submitted.
    \item\label{p:profiles} Job submission is confined to per-supercomputer profiles, so the engine is workload-manager agnostic.
    Once a campaign's tasks avoid values specific to one supercomputer, moving it to another means writing profiles and setting parameters, not editing tasks.
    \item\label{p:records} Every run writes a provenance record next to its output (environment, hardware, parameters, git state) and marks its outcome, so failed runs can be listed and rerun on their own.
    \item\label{p:engine} The engine is a set of Bash scripts that is copied into each campaign and needs no installation.
    \item\label{p:any} Tasks can call anything, such as compilers, MPI launchers, Julia, gnuplot, or Python.
\end{enumerate}

\Thinflow targets researchers who already write job scripts for recurring benchmark sweeps, move them between supercomputers, and want to publish an artifact.
Workflows whose shape depends on intermediate results, large ensembles, data-heavy pipelines, and the construction of software environments are out of scope.
The tools in Section~\ref{sec:related} serve them better.
\Thinflow aims at a small footprint: its user interface consists of five file names, ten reserved keys, a brace syntax for run IDs, and 14 command-line flags.

This paper makes the following contributions:
\begin{enumerate}[label=\textbf{C\arabic*}, ref=C\arabic*, leftmargin=*]
    \item\label{c:model} A campaign model of directories with inherited settings and a plain-text manifest of the planned execution (Section~\ref{sec:design}).
    \item\label{c:impl} A Bash implementation with per-supercomputer profiles and per-run provenance records that ships inside the campaign it runs (Section~\ref{sec:impl}).
    \item\label{c:case} A case study in which the same 37 task files ran a three-variant strong-scaling campaign of 455 benchmark runs on three Slurm partitions of two supercomputers, with the problems we met when moving it and where each fix belongs (Section~\ref{sec:eval}).
\end{enumerate}

\section{Background and Related Work}
\label{sec:related}

\subsection{Reproducible Benchmarking}
Hoefler and Belli examined 120 papers from SC, HPDC, and PPoPP and found that most describe their hardware but not their software environment, and that few report any measure of variation.
Their guidelines presuppose that the path from configuration to result is explicit~\cite{hoefler2015benchmarking}.
Koskela et al.\ turn such principles into tooling, built on ReFrame and Spack, that records the build configuration, figure of merit, and platform of every run~\cite{koskela2023principles}.
Antunes and Hill list HPC-specific obstacles to reproducibility such as parallel nondeterminism, compiler optimization, and hardware heterogeneity~\cite{antunes2024reproducibility}.
Diagnosing any of them requires a record of what ran, in which order, and with which inputs.

Pearce et al.\ distinguish three operations on a benchmarking experiment~\cite{pearce2025rrr}.
To \emph{repeat} it means to rerun it on the same supercomputer with the same hardware, software stack, and application, for example by another user or at a later time.
To \emph{replicate} it means to run it on the same hardware with a possibly different software stack, such as other compiler or library versions.
To \emph{reproduce} it means to run it on a supercomputer with different hardware, keeping the experimental parameters as similar as possible and recording every purposeful difference.
We use these terms throughout.

Venues increasingly expect a reusable artifact.
ACM awards badges for available, functional, and reusable artifacts and for reproduced results~\cite{acmbadging}, and SC asks for an artifact description and offers an optional evaluation stage~\cite{sc26adae}, a process refined over several years~\cite{malik2022sc21,pallez2025sc24}.
How far authors embrace these initiatives varies~\cite{sirvent2025reproducibility}.
Frachtenberg finds an artifact-sharing rate near 30\% and about one in five shared artifacts inaccessible four years after publication~\cite{frachtenberg2022artifacts}.
Guilloteau et al.\ find that practice in almost 300 papers from 2023 does not address artifact longevity and propose a badge based on source code, experimental setup, and software environment~\cite{guilloteau2024longevity}.
\emph{\Thinflow writes a provenance record for every run (Section~\ref{sec:impl-records}) and ships inside the campaign, which can thus serve as the artifact.}

\subsection{Benchmarking Frameworks}
\label{sec:related-frameworks}
JUBE describes a benchmark in an XML or YAML file with parameter sets and steps that run shell commands~\cite{steinbach2026jube}.
Platform files hold supercomputer-specific settings and job-script templates, each combination of parameters runs as a workpackage in its own numbered directory with a completion marker, and patterns and analysers turn the output into result tables.
In ReFrame, tests are Python classes, and a site configuration describes supercomputers, partitions, workload managers, launchers, and programming environments, so that one test runs on several supercomputers~\cite{karakasis2020reframe}.
Benchpark captures benchmark and system specifications so that experiments can be shared and replayed across sites~\cite{pearce2023benchpark}.
Ramble renders them into workspaces, provides the software through Spack or other package managers, and records an inventory with hashes of inputs, software, and templates~\cite{jacobsen2023ramble}.
Benchpark's analysis step\footnote{\url{https://software.llnl.gov/benchpark/benchpark-analyze.html}} produces scaling figures with Thicket, which analyzes the performance data of many runs across scales and supercomputers~\cite{brink2023thicket}.
Taskblaster defines workflows in Python, stores each task's output in a directory tree, and keeps the task graph in an SQLite registry file, without a database server~\cite{larsen2025taskblaster}.
Maestro\footnote{\url{https://github.com/LLNL/maestrowf}} specifies studies as YAML files with shell steps and parameters and runs every parameter combination in its own workspace.
Merlin extends it to large ensembles~\cite{peterson2022merlin}.
The signac framework keeps the state of a parameter study as one directory per parameter combination~\cite{adorf2018signac}, and the Popper convention applies DevOps practices such as version control and automated pipelines to experiments in systems research~\cite{jimenez2017popper}.
Badwaik and Herten describe the structure of a benchmark: benchmark, system, and instrumentation definitions are resolved into a space of concrete benchmark specifications, whose results feed the generated report~\cite{badwaik2026unified}.
\emph{\Thinflow fills this structure with directories and plain shell files instead of an XML, YAML, or Python description (Section~\ref{sec:design-structure}), and Table~\ref{tab:concepts} relates it to JUBE, ReFrame, and Benchpark with Ramble.}

\begin{table*}[t]
    \caption{
        Concepts of a benchmarking campaign in three established frameworks and in \Thinflow, as described in each tool's documentation (October 2026).
        The comparison is qualitative, not measured.
    }
    \label{tab:concepts}
    \tablestyle
    \begin{tabularx}{\textwidth}{H{2.3cm}W{0.92}W{0.95}W{1.03}W{1.10}}
        \toprule
        & \multicolumn{3}{l}{\textbf{Established frameworks}} & \multicolumn{1}{l}{\textbf{This paper}} \\
        \cmidrule(r){2-4} \cmidrule{5-5}
        Concept & \textbf{JUBE} & \textbf{ReFrame} & \textbf{Benchpark with Ramble} & \textbf{\Thinflow} \\
        \midrule
        Experiment definition & XML or YAML file with parameter sets and steps & Python test classes & Python experiment and system specifications that generate a Ramble workspace & directory tree of shell files with inherited settings \\
        \shaded Parameter sweep & comma-separated values; one workpackage per combination & \texttt{parameter()} dimensions & variables, vectors, and matrices & brace expansion of run IDs in \texttt{TASK\_RUNS} \\
        Supercomputer and job settings & platform files with job-script templates & site configuration: partitions, workload managers, launchers & system specification: hardware, compilers, MPI, workload manager, launcher & per-supercomputer profile scripts \\
        \shaded Software environment & commands in steps & programming environments & Ramble package managers (Spack, pip, EESSI, modules, user-managed) & modules in profiles or environment scripts, or a container image optionally checked against its definition file \\
        Figures of merit & patterns and analysers & sanity and performance functions & figures of merit and success criteria & user scripts in tasks \\
        \shaded Analysis and plots & result tables; plots via user steps & performance logs, run reports & scaling figures via Thicket & plot tasks running user scripts \\
        Record per run & configuration copy, logs, stdout and stderr & performance log, JSON report & inventory with hashes of inputs, software, templates & environment, hardware, git state, overrides, outcome \\
        \shaded State on disk & numbered directories with done markers & stage and output directories, report database & workspace directories & run directories, state markers, text manifest \\
        Installed beforehand & Python~3 (PyYAML for YAML input) & Python package & Python; Benchpark clones Spack and Ramble & Bash 4.4 and GNU tools; engine copied into each campaign \\
        \bottomrule
    \end{tabularx}
\end{table*}

\subsection{Workflow Systems, Sweeps, and Containers}
General workflow systems, which Ferreira da Silva et al.\ characterize in a common space~\cite{ferreiradasilva2017characterization}, also separate workload-manager specifics from the workflow definition.
Parsl wraps \texttt{sbatch}, \texttt{squeue}, and \texttt{srun} behind provider and launcher abstractions~\cite{babuji2019parsl}, Snakemake passes per-rule resources to Slurm or PBS and keeps plotting rules in the same graph as the measurements~\cite{molder2021snakemake}, Balsam and FireWorks keep task queues in a database~\cite{salim2019balsam,jain2015fireworks}, and Pegasus maps an abstract workflow onto an executable plan~\cite{deelman2015pegasus}.
Sweep tools range from OACIS, with a web server and MongoDB~\cite{murase2017oacis}, to parasweep and Spinner, which generate one input per parameter set~\cite{bach2021parasweep,ceccato2024spinner}.
PSI/J offers a small job API for several workload managers~\cite{hategan2023psij}.
Container runtimes such as Singularity, Charliecloud, Shifter, and Sarus provide portable software stacks on supercomputers~\cite{kurtzer2017singularity,priedhorsky2016charliecloud,gerhardt2017shifter,benedicic2019sarus,sochat2017registry}, and Spack builds software stacks from source~\cite{gamblin2015spack}.
\emph{\Thinflow needs no server or database, keeps its state in run directories (Section~\ref{sec:impl-operating}), and builds no software itself: a task file can activate a Spack environment, and a run can execute in a container.}

\section{\Thinflow Design}
\label{sec:design}
\begin{figure*}[t]
\begin{lstlisting}
tasks/
|- run_meta.sh            RUN_WORKLOAD_MANAGER=$TF_WM_BASE.sh     /*profile, if TF_WM_BASE is set*/
|- build/
|  |- task_meta.sh        TASK_RUNS="cpu"
|  |- opt/run.sh          $CC -O3 $ARCH -fopenmp matmul.c -o matmul /*first OpenMP-capable compiler*/
|  |  |- cpu/             // C/OpenMP binary
|  |- mpi/run.sh          mpicc -O3 $ARCH matmul_mpi.c -o matmul
|  |  |- cpu/             // C/MPI binary
|  |- mpijulia/run.sh     julia --project=... -e 'using MPI ...' /*check MPI.jl uses the supercomputer's MPI*/
|     |- run_meta.sh      RUN_WORKLOAD_MANAGER=$TF_WM_BASE-mpi.sh
|- experiment/
|  |- task_meta.sh        TASK_RUNS="{t1,t8,...,t192}-run{1:5}"   /*from SCALING_LEVELS, SCALING_REPS*/
|  |- MatMul_SingleNode/
|  |  |- task_meta.sh     ROUTINE=MatMul; N=${N:-4096}
|  |  |- run_env.sh       P=/*parallelism, parsed from RUN_ID*/
|  |  |- omp/
|  |  |  |- run_env.sh    run_benchmark() { OMP_NUM_THREADS=$P .../build/opt/cpu/matmul $N; }
|  |  |  |- run_deps.sh   RUN_DEPENDENCIES=(tasks/build/opt:cpu)
|  |  |  |- run.sh        run_benchmark > stdout
|  |  |- mpi/             (analogous; mpirun -np $P, or srun if the profile sets MPI_LAUNCHER=srun)
|  |  |- mpijulia/        (analogous; the same launch for julia matmul_mpi.jl $N)
|  |- MatMul_MultipleNodes/
|     |- task_meta.sh     N=${N:-8192}; TASK_RUNS="{n1,n2,...,n16}-run{1:5}" /*from TF_MN_NODES*/
|     |- mpi/ mpijulia/   (as above; run_meta.sh selects $TF_WM_BASE-<n>n.sh per node count)
|- plot_SingleNode/
|  |- task_meta.sh        TASK_RUNS="output"
|  |- run_deps.sh         RUN_DEPENDENCIES=(tasks/.../omp tasks/.../mpi tasks/.../mpijulia)
|  |- run.sh              sh aggregate.sh; gnuplot plot.gp
|- plot_MultipleNodes/    (analogous; renders the multi-node figure)
\end{lstlisting}
\caption{
    The campaign of the case study in Section~\ref{sec:eval} as a directory tree (left: files and directories, right: their simplified content).
    The \texttt{build} group compiles the variants, the \texttt{experiment} group sweeps each variant over the parallelism levels with five repetitions, and two \texttt{plot} tasks render the figures.
    Lists marked \texttt{...} are abbreviated.
    Directories marked \emph{(analogous)} repeat the \texttt{omp} leaf, differing only in the part shown.
    The sweep, the problem size $N$, and the parallelism $P$ are declared once and inherited.
    The values used in Section~\ref{sec:eval} are set on the command line.
    Only \texttt{run\_meta.sh} files name a profile, and no task file submits jobs.
    The \texttt{run\_benchmark} functions in \texttt{run\_env.sh} call \texttt{mpirun}, or \texttt{srun} if the profile sets \texttt{MPI\_LAUNCHER=srun}.
}
\label{fig:mmtree}
\end{figure*}

\Thinflow's campaign model (\ref{c:model}) treats a campaign as units of work that may each run many times, with dependencies between those runs.
Besides properties \ref{p:tree} to \ref{p:any}, the design follows two goals: it imposes no coding or experimentation style on the work being run, and it is additive, placing files alongside the existing code and leaving that code untouched.

\subsection{Tasks and Runs}
\label{sec:design-primitives}
Any directory that contains a \texttt{run.sh} file is a \emph{task}.
A task usually executes as several \emph{runs}, for example once per parallelism level and repetition.
Each run has an ID that is unique within its task and executes the task's \texttt{run.sh} in its own output directory named after that ID.
Figure~\ref{fig:mmtree} shows a complete campaign laid out this way: the three-variant matrix-multiply scaling study of Section~\ref{sec:eval}.

\subsection{Settings, Run Dependencies, and Environment}
\label{sec:design-aspects}
Tasks and runs carry settings as \texttt{KEY=VALUE} pairs, task settings in \texttt{task\_meta.sh} and run settings in \texttt{run\_meta.sh}.
Ten keys are reserved.
The central one is \texttt{TASK\_RUNS}, with which a task declares how many runs it has and which IDs they receive.
The ID specification uses a brace syntax in which several brace groups expand as a Cartesian product: \texttt{"\{t1,t8,t16\}-run\{1:5\}"} yields five repetitions at each of three parallelism levels, and \texttt{"\{cpu,gpu\}-\allowbreak run\{1:10\}"} yields ten repetitions on each of two devices.
Figure~\ref{fig:mmsweep} shows how the case study declares its sweep.

\begin{figure}[t]
\begin{lstlisting}
# experiment/task_meta.sh  -- declare the sweep
_reps=${SCALING_REPS:-3}
if [ -n "${SCALING_LEVELS:-}" ]; then
  _list=...   # t1,t8,... from SCALING_LEVELS
else
  _list=...   # powers of two up to the core count
fi
TASK_RUNS="{$_list}-run{1:$_reps}"
# 192 cores: {t1,t8,t16,t32,t64,t96,t128,t192}-run{1:5}

# experiment/MatMul_SingleNode/task_meta.sh  -- shared
ROUTINE=MatMul
N=${N:-4096}

# .../MatMul_SingleNode/omp/run.sh  -- the work
run_benchmark > stdout  # from the variant's run_env.sh
\end{lstlisting}
\caption{
    Declaring the single-node sweep of Figure~\ref{fig:mmtree} (abbreviated).
    With the values used on the 192-core partition, \texttt{TASK\_RUNS} expands to eight parallelism levels $\times$ five repetitions, 40 runs per variant.
    The problem size is set once and inherited, and \texttt{run.sh} is the same for every run.
}
\label{fig:mmsweep}
\end{figure}

A run can name a container image (\texttt{RUN\_CONTAINER}) in which its \texttt{run.sh} executes.
It can also name the definition file from which the image must have been built (\texttt{RUN\_\allowbreak CONTAINER\_\allowbreak DEF}).
A run's \texttt{run\_meta.sh} may likewise choose the profile that submits it (\texttt{RUN\_\allowbreak WORKLOAD\_\allowbreak MANAGER}, Section~\ref{sec:impl}).

\emph{Run dependencies} are declared in \texttt{run\_deps.sh}, whose array \texttt{RUN\_\allowbreak DEPENDENCIES} names other runs in the same brace syntax.
A run starts only once all of its dependencies have succeeded.
With the Slurm profiles, however, each job depends on all jobs of the previous stage, so a failed run cancels all later stages.
\Thinflow does not manage \emph{software dependencies}: compilers and libraries come from the supercomputer's modules loaded by a profile, from a container image, or from an environment script (\texttt{run\_env.sh}), for example one that activates a Spack environment.

The environment of a run is assembled from its environment scripts, which define variables and helper functions such as the \texttt{run\_benchmark} function of Figure~\ref{fig:mmtree}.
Command-line \texttt{KEY=VALUE} overrides on \texttt{run\_\allowbreak tasks.sh}, such as \texttt{N=8192}, are recorded in the manifest and applied when a run's environment is built via \texttt{run\_\allowbreak env.sh}.
A variable such as the problem size $N$ thus reaches every selected run, and each run records it (Section~\ref{sec:impl-records}).
We call \texttt{run.sh} and these settings, dependency, and environment files the \emph{task files} of a campaign.

\subsection{Hierarchy and Inheritance}
\label{sec:design-hierarchy}
Tasks are organized in a directory tree for grouping, so that, for example, all compilation tasks sit under one \texttt{build/} directory.
Settings, dependencies, and environment may be defined at every level of that tree, and values at a lower level override values at a higher level (\ref{p:tree}).
A parent directory thus sets defaults that individual tasks and runs refine, which keeps each leaf small: a benchmark group sets the routine and problem size once, and each variant directory adds only what distinguishes it.

\subsection{Three-Phase Execution Model}
\label{sec:design-execution}
A campaign is launched with \texttt{./run\_tasks.sh}, optionally followed by the tasks to run (default: all of \texttt{tasks/}).
Execution proceeds in three phases.
Figure~\ref{fig:phases} lists what each phase reads and writes.

\begin{figure}[t]
\centering
\begin{tikzpicture}[
  font=\footnotesize,
  box/.style={draw, rounded corners, align=left, inner sep=4pt, text width=0.92\linewidth},
  wmbox/.style={draw, dashed, rounded corners, align=left, inner sep=4pt, text width=0.92\linewidth},
  arr/.style={-{Stealth[length=2mm]}, thick},
  node distance=4mm,
]
\node[box] (p1) {\textbf{1. Task Meta Phase}\\
  \emph{reads} \texttt{task\_meta.sh} at every level of each task path\\
  \emph{derives} the runs of each task from \texttt{TASK\_RUNS}};
\node[box, below=of p1] (p2) {\textbf{2. Run Meta Phase}\\
  \emph{reads} \texttt{run\_meta.sh} and \texttt{run\_deps.sh} per run\\
  \emph{groups} runs into stages of mutually independent runs\\
  \emph{writes} the \emph{manifest} (printed by \texttt{-{}-dry-run})};
\node[wmbox, below=of p2] (wm) {\textbf{Profile (workload-manager script), per stage}\\
  \emph{reads} the manifest and hands each group of runs to the workload manager (with Slurm: one array job, chained with \texttt{afterok}), or runs it locally};
\node[box, below=of wm] (p3) {\textbf{3. Run Phase}, per run in its own directory\\
  \emph{clears} the directory; \emph{writes} \texttt{.run\_begin}\\
  \emph{sources} \texttt{task\_meta.sh} and \texttt{run\_meta.sh}\\
  enters the container if set (image optionally checked against a definition file)\\
  \emph{sources} \texttt{run\_env.sh}; \emph{writes} \texttt{.run\_metadata}\\
  \emph{sources} \texttt{run.sh}; \emph{writes} \texttt{.run\_success} or \texttt{.run\_failed}};
\draw[arr] (p1) -- (p2);
\draw[arr] (p2) -- (wm);
\draw[arr] (wm) -- (p3);
\end{tikzpicture}
\caption{
  The three-phase execution model and the files each phase reads and writes.
  At each step, the files of every level of the task's path are sourced from the root down.
  The manifest is the hand-off between planning and execution, and workload-manager-specific code is confined to the profiles and a helper they call.
}
\label{fig:phases}
\end{figure}

The \emph{Task Meta Phase} derives the complete list of runs, and the \emph{Run Meta Phase} groups them into stages of mutually independent runs and writes the manifest.
Because the stages already encode independence, a Slurm profile can submit a whole stage as concurrent array jobs without further analysis, while the local profile runs it sequentially.
The \emph{Run Phase} then executes each run in its own directory.

\subsection{Worked Example}
\label{sec:design-example}
In the campaign of Figure~\ref{fig:mmtree}, the \texttt{build} group first compiles the C/OpenMP and C/MPI variants and checks the Julia project, with one \texttt{cpu} run per variant.
The \texttt{experiment} group combines a shared problem definition with a sweep: \texttt{TASK\_RUNS} expands into the parallelism levels times the repetitions, the problem size $N$ is inherited, and each variant leaf adds only its \texttt{run\_benchmark} and its build dependency.
Each experiment run depends on its variant's build run, so the Run Meta Phase places all builds in an earlier stage than any benchmark.
The two \texttt{plot} tasks depend on the experiment runs of their variants, which places them in the final stage, where they aggregate the measurements and render the figures.

\subsection{Isolation and Reruns}
\label{sec:design-isolation}
Every run executes in its own directory, so concurrent runs of the same task cannot collide, and a failed run leaves the directories of other runs untouched.
Rerunning a run clears its directory and executes it again.
The resulting directory tree can be read without the tool: its structure shows which runs exist, and each run directory holds the run's record and output.

\subsection{\Thinflow in a Common Benchmark Structure}
\label{sec:design-structure}
Table~\ref{tab:structure} places \Thinflow's elements in the benchmark structure of Badwaik and Herten (Section~\ref{sec:related-frameworks}).
The manifest makes the generated space explicit as a file that can be read before anything runs.

\begin{table}[t]
  \caption{\Thinflow's elements in the benchmark structure of Badwaik and Herten~\cite{badwaik2026unified}.}
  \label{tab:structure}
  \tablestyle
  \begin{tabularx}{\columnwidth}{H{2.75cm}Y}
    \toprule
    Element & \textbf{\Thinflow} \\
    \midrule
    Benchmark definitions & task directories with \texttt{task\_meta.sh} and \texttt{run.sh} \\
    \shaded System definitions & per-supercomputer profiles, chosen in \texttt{run\_meta.sh} \\
    Instrumentation definitions & launch wrappers in \texttt{run\_env.sh}; metric extraction in plot tasks \\
    \shaded Resolution & inheritance along the tree and \texttt{TASK\_RUNS} expansion (Task and Run Meta Phases) \\
    Concrete benchmark specifications & the manifest (\texttt{-{}-dry-run}) \\
    \shaded Results & run directories with output and \texttt{.run\_metadata} \\
    Report generation & plot tasks running user scripts \\
    \bottomrule
  \end{tabularx}
\end{table}

\section{Implementation}
\label{sec:impl}

The implementation (\ref{c:impl}) comprises about 3,100 lines of Bash (2,600 without comments and blank lines), including a local profile and a container manager, and is copied into each campaign (\ref{p:engine}).
It needs Bash 4.4 or newer, because it uses associative arrays and name references and expands possibly empty arrays under \texttt{set -u}, and it expects the GNU tools of Linux.
Task files are sourced by the same Bash, while the programs that tasks call, such as the Julia benchmark and the gnuplot scripts of the case study, can be written in any language (\ref{p:any}).
The single entry point, \texttt{run\_tasks.sh}, walks the task tree, sources the settings and dependency files of every level from the root down so that lower levels override higher ones, builds the run dependency graph, and partitions it into stages.
The result is a plain-text manifest (\ref{p:tree}, Figure~\ref{fig:manifest}) with one \texttt{JOB} block per group of runs that share a stage, a profile, and a workload name (the job name passed to the workload manager).
Each block names its stage, its profile, the blocks it depends on (all blocks of the previous stage), and an indexed list of runs.
\texttt{-{}-dry-run} prints this manifest, so the planned execution can be checked before a single job is submitted.

\begin{figure}[t]
\begin{lstlisting}
JOB               0
STAGE             0
WORKLOAD_NAME     run_tasks: build/mpijulia
WORKLOAD_MANAGER  .../<name>-mpi.sh
DEPENDS
0  cpu      tasks/build/mpijulia
---
JOB 1  (stage 0, build/mpi, <name>.sh)
---
JOB               2
STAGE             1
WORKLOAD_NAME     run_tasks: .../MatMul_MultipleNodes/*
WORKLOAD_MANAGER  .../<name>-1n.sh
DEPENDS           0,1
0  n1-run1  tasks/experiment/MatMul_MultipleNodes/mpi
   ...      (10 runs: 2 variants x 5 repetitions)
---
JOB 3 to JOB 6  (stage 1, <name>-2n.sh to -16n.sh)
---
JOB               7
STAGE             2
WORKLOAD_NAME     run_tasks: plot_MultipleNodes
WORKLOAD_MANAGER  .../<name>.sh
DEPENDS           2,3,4,5,6
0  output   tasks/plot_MultipleNodes
\end{lstlisting}
\caption{
    The manifest of the multi-node invocation of the case study, abbreviated\AnonOnly{, with supercomputer names elided}.
    The file is tab-separated.
    Stage~0 builds, stage~1 benchmarks with one job per node count, stage~2 plots.
    The stages and dependencies are derived from the task tree.
}
\label{fig:manifest}
\end{figure}

A profile (\ref{p:profiles}) is a workload-manager script that receives the manifest and a stage number, and executes the job blocks of that stage that name it.
The local profile (60 lines) runs them on the current machine, one run after another.
For Slurm, a shared helper submits each job block as one array job and translates its dependencies into \texttt{afterok} dependencies on the Slurm job IDs of those blocks.
No other engine code is Slurm-specific, so supporting another workload manager requires new profiles and no change to the engine.

A container manager script, Apptainer by default, runs \texttt{run.sh} inside the image named by \texttt{RUN\_\allowbreak CONTAINER}, so the container runtime is exchangeable like the profile.
If \texttt{RUN\_\allowbreak CONTAINER\_\allowbreak DEF} names a definition file, the run first compares it with the definition embedded in the image and refuses to start if they differ, unless \texttt{-{}-skip-\allowbreak verify-\allowbreak def} is given.
The case study does not use containers.

\subsection{What a Run Records}
\label{sec:impl-records}
Before \texttt{run.sh} starts, each run writes a \texttt{.run\_metadata} file into its directory (\ref{p:records}, Figure~\ref{fig:metadata}).
This file is the run's provenance record, the account of how its result was produced.
It starts with the engine's version string and the completion times of the runs it depends on.
It then records the git commit of the campaign, if the campaign is a git repository, and the names of files with uncommitted changes in \texttt{assets/} (the campaign's sources and inputs) and in the dependency runs.
The rest holds the \texttt{KEY=VALUE} overrides given on the command line, the reserved run settings such as the profile and container, the result of the container check, the complete environment, and the hardware as reported by \texttt{lscpu}, \texttt{free}, and \texttt{nvidia-smi}, or ``N/A'' where these Linux tools are missing.
Marker files record the start time and the outcome (\texttt{.run\_begin}, \texttt{.run\_success}, \texttt{.run\_failed}).

\begin{figure}[t]
\begin{lstlisting}
=== git ===
git_commit=2571484e9c7094fe670327d6d85f8942a4d9e52a
--- git status porcelain: $ASSETS ---
 M assets/julia/Manifest.toml
?? assets/julia/LocalPreferences.toml
git_uncommitted_summary=dirty
=== framework variables ===
RUN_WORKLOAD_MANAGER=.../<name>-mpi.sh
RUN_ID=t1-run1
=== environment ===
LOADEDMODULES=...:compiler/GCC/14.2.0:...
  :mpi/OpenMPI/5.0.7-GCC-14.2.0:...
  :lang/JuliaHPC/1.12.6-foss-2025a
N=8192
SCALING_LEVELS=1 8 16 32 64 96 128 192
SCALING_REPS=5
=== hardware ===
cores: 192
Model name:  AMD EPYC 9655 96-Core Processor
Socket(s):   2
\end{lstlisting}
\caption{
    Excerpt of the \texttt{.run\_metadata} of one Julia/MPI run of the case study (shortened\AnonOnly{, supercomputer names elided}).
    The record shows the modules that were loaded, and it flags two Julia files as uncommitted changes, the updated package manifest and the preferences written on the login node (Section~\ref{sec:eval-lessons}).
}
\label{fig:metadata}
\end{figure}

\subsection{Operating a Campaign}
\label{sec:impl-operating}
The same entry point offers flags for developing, inspecting, and recovering a campaign.
Naming task paths runs only those subtrees.
A suffix on a task path, such as \texttt{:t8-run\{1:5\}} in the syntax of \texttt{TASK\_RUNS}, selects individual runs.
\texttt{KEY=VALUE} overrides apply to the task paths that follow them, for example to rerun one parallelism level at a larger problem size.
\texttt{-{}-include-deps} pulls the runs a target depends on into the same invocation, so asking for a plot also builds and benchmarks what it needs.
\texttt{-{}-skip-succeeded} reuses runs that already succeeded, and \texttt{-{}-clean} removes run directories.

Because each run records its outcome in its own directory, the state of a campaign can be read from the filesystem.
\texttt{run\_tasks.sh -{}-status} prints a table of \texttt{SUCCESS}, \texttt{FAILED}, \texttt{RUNNING}, and \texttt{PENDING} runs (Figure~\ref{fig:status}).
Every launch also leaves a \texttt{status.sh} next to its manifest that prints this table for exactly the launched runs, without requiring their task paths to be listed again.
After a partial failure, the user fixes the cause and invokes the same command again with \texttt{-{}-skip-succeeded}, which executes only the runs without a success marker, including those of cancelled stages.

\begin{figure}[t]
\begin{lstlisting}
$ ./run_tasks.sh --status tasks/experiment/MatMul_SingleNode
JOB/IDX  RUN      PATH                            STATUS
0/0      t1-run1  .../MatMul_SingleNode/omp       SUCCESS
0/1      t8-run2  .../MatMul_SingleNode/omp       FAILED
1/0      t16-run1 .../MatMul_SingleNode/mpijulia  SUCCESS

# fix the cause, then run only what failed:
$ ./run_tasks.sh --skip-succeeded \
      tasks/experiment/MatMul_SingleNode
\end{lstlisting}
\caption{
    Example of the run state read from the per-run marker files (columns abbreviated): \texttt{-{}-status} lists each run's outcome, and \texttt{-{}-skip-succeeded} executes only the runs without a success marker.
}
\label{fig:status}
\end{figure}

\subsection{Regression Tests}
\label{sec:impl-tests}
A regression suite of 19 mini-repositories, each a miniature campaign with its own expected outputs, holds 116 test cases that assert the exit code, standard output, and file effects of every invocation.
It covers argument parsing, task and run resolution, dependencies, the manifest, execution with the local profile, and the status table.

\section{Case Study: Scaling Campaign}
\label{sec:eval}

The case study (\ref{c:case}) splits the question of Section~\ref{sec:intro} into four parts:
\begin{itemize}
  \item whether identical task files can drive a complete campaign, from builds to plots, on Slurm partitions of different size and software stack
  \item what such a campaign costs to write
  \item which problems arise when it moves between supercomputers, and where their fixes belong
  \item whether the run records show what differed between the partitions
\end{itemize}
The benchmark serves these questions and is not a contribution in itself.
We do not rank programming models.

\subsection{Campaign}
\label{sec:eval-campaign}
\begin{figure*}[t]
\centering
\begin{tikzpicture}[
  font=\footnotesize,
  >={Stealth[length=1.6mm]},
  cmd/.style={align=center, inner sep=2pt},
  job/.style={draw, rounded corners, align=center, inner sep=3pt},
  jobs/.style={job},
  head/.style={font=\small\bfseries, align=center},
]
\node[job, fill=blue!7, text width=24mm] (bopt) {{\small C/OpenMP}\\\texttt{gcc -fopenmp}\\\textit{1 node}};
\node[job, fill=blue!7, text width=24mm, below=10.5mm of bopt] (bmpi) {{\small C/MPI}\\\texttt{mpicc}\\\textit{1 node}};
\node[job, fill=blue!7, text width=24mm, below=10.5mm of bmpi] (bdag) {{\small Julia/MPI}\\\texttt{julia} + MPI.jl\\\textit{1 node}};
\node[cmd, left=7mm of bmpi] (cmd) {\texttt{./run\_tasks.sh}};
\node[anchor=north west, font=\footnotesize\itshape] at (cmd.south west) {login node};
\node[job,  fill=green!9, text width=44mm, right=9mm of bopt] (eomp) {{\small C/OpenMP runs}\\\textit{1 job, 1 node}\\\textit{threads $1,8,\ldots,P$}\\\textit{$R$ reps each}};
\node[jobs, fill=green!9, text width=44mm, right=9mm of bmpi] (empi) {{\small C/MPI runs}\\\textit{ranks $1,8,\ldots,P$: one 1-node job}\\\textit{nodes $1,2,4,\ldots$: one job each}\\\textit{shared with Julia/MPI; $R$ reps each}};
\node[jobs, fill=green!9, text width=44mm, right=9mm of bdag] (edag) {{\small Julia/MPI runs}\\\textit{ranks $1,8,\ldots,P$: one 1-node job}\\\textit{nodes $1,2,4,\ldots$: one job each}\\\textit{shared with C/MPI; $R$ reps each}};
\node[job, fill=orange!14, text width=28mm, right=11mm of eomp] (pltSN) {\texttt{plot\_SingleNode}\\\textit{omp, mpi, mpijulia}\\\textit{single node}};
\node[job, fill=orange!14, text width=28mm, right=11mm of edag] (pltMN) {\texttt{plot\_MultipleNodes}\\\textit{mpi, mpijulia}\\\textit{per node count}};
\node[minimum width=13mm, minimum height=11mm] (outSN) at ([xshift=14mm]pltSN.east) {};
\node[minimum width=13mm, minimum height=11mm] (outMN) at ([xshift=14mm]pltMN.east) {};
\foreach \o in {outSN,outMN}{
  \coordinate (oo) at ([xshift=1.5mm,yshift=1.5mm]\o.south west);
  \draw[gray, ->] (oo) -- ++(0,7mm);
  \draw[gray, ->] (oo) -- ++(10mm,0);
  \draw[blue!70!black, very thick] (oo)++(0.5mm,0.5mm) -- ++(2.4mm,2mm) -- ++(2.4mm,1.2mm) -- ++(3.2mm,2.8mm);
}
\foreach \b in {bopt,bmpi,bdag}{ \draw[->] (cmd) -- (\b); }
\draw[->] (bopt) -- (eomp);
\draw[->] (bmpi) -- (empi);
\draw[->] (bdag) -- (edag);
\foreach \dx in {3.6,5.4}{
  \draw[->, gray] (bmpi.east) -- ([xshift=\dx mm, yshift=\dx mm]empi.north west);
  \draw[->, gray] (bdag.east) -- ([xshift=\dx mm, yshift=\dx mm]edag.north west);
}
\draw[->] (eomp.east) -- (pltSN.west);
\draw[->] (empi.east) -- (pltSN.west);
\draw[->] (edag.east) -- (pltSN.west);
\draw[->] (empi.east) -- (pltMN.west);
\draw[->] (edag.east) -- (pltMN.west);
\draw[->, dashed] (pltSN.east) -- (outSN);
\draw[->, dashed] (pltMN.east) -- (outMN);
\node[head, above=3.5mm of bopt] (h1) {Stage 0:\\Build};
\node[head] at (cmd  |- h1) {Login node:\\Submit};
\node[head] at (eomp |- h1) {Stage 1:\\Experiment};
\node[head] at (pltSN |- h1) {Stage 2:\\Plot};
\node[head] at (outSN  |- h1) {Result:\\Figures};
\node[job, fill=white, minimum width=6mm, minimum height=4mm, anchor=north west] (lg1) at ([xshift=1mm, yshift=-7mm]pltSN.south west) {};
\node[right=6.5mm of lg1, anchor=west] {$=$ runs of one variant};
\node[jobs, fill=white, minimum width=6mm, minimum height=4mm, anchor=north west] (lg2) at ([yshift=-9mm]lg1.south west) {};
\node[right=6.5mm of lg2, anchor=west] {$=$ one job per node count};
\begin{scope}[on background layer]
  \foreach \nd in {empi,edag,lg2}{
    \foreach \dx in {5.4,3.6,1.8}{
      \draw[fill=white, draw=black!55, rounded corners]
        ([xshift=\dx mm, yshift=\dx mm]\nd.south west) rectangle
        ([xshift=\dx mm, yshift=\dx mm]\nd.north east);
    }
  }
\end{scope}
\end{tikzpicture}
\caption{
  How the campaign of Figure~\ref{fig:mmtree} maps onto Slurm jobs on one partition.
  The two invocations on the login node, one per plot task, expand into the run dependency graph with three stages, as in the manifest of Figure~\ref{fig:manifest}: three build runs (stage~0), the scaling runs of each variant (stage~1), and two plot runs (stage~2).
  A box groups the runs of one variant in one stage, and a stack of boxes denotes one job per node count.
  Runs of a stage that share a profile are submitted together as one Slurm array job (Figure~\ref{fig:manifest}).
  C/OpenMP runs on one node only and feeds only the single-node plot.
  The two MPI variants share a single-node job and one right-sized job per node count, and both feed the multi-node plot.}
\label{fig:campaign}
\end{figure*}

The benchmark computes the dense matrix product $C = A \times B$ of size $N \times N$ in three variants.
All variants use the same na\"ive $O(N^3)$ arithmetic in $ikj$ loop order and no BLAS, so no vendor library hides differences between the programming models.
The C/OpenMP variant (\texttt{omp} in the task tree) is a C kernel parallelized with OpenMP.
The C/MPI variant (\texttt{mpi}) is the same C kernel under MPI, distributing the rows of $A$ across ranks, broadcasting $B$, and gathering the result blocks at the root.
The Julia/MPI variant (\texttt{mpijulia}) implements the same MPI algorithm in Julia via MPI.jl\footnote{\url{https://github.com/JuliaParallel/MPI.jl}}.
It adds a second toolchain: a Julia project that must be instantiated and bound to the supercomputer's MPI before compute nodes without network access can use it.

Threads and ranks are placed compactly, one per physical core.
One untimed warm-up per run keeps just-in-time compilation and cold-start effects out of the measurement.
The timed region covers the parallel multiply and, for the MPI variants, the gather.
Every variant generates $A$ and $B$ from the same deterministic formulas and prints the same checksum, which serves as a correctness check across variants.

The campaign is the task tree of Figure~\ref{fig:mmtree}: 217 non-comment lines of shell in 37 files, plus 377 non-comment lines of aggregation and gnuplot scripts in \texttt{assets/plots}.
It adds no code to the benchmark sources.
Each variant is swept over parallelism levels within a node and, for the two MPI variants, over node counts, with five repetitions per point.
Two plot tasks aggregate the metric line that every run prints, with a small shell and awk script, and render the figures with gnuplot.
For each point, the figures show the mean execution time and throughput with their standard deviation over the repetitions, and the speedup $t_1/t_P$.

On each partition, after the Julia project had been set up on the login node (Section~\ref{sec:eval-lessons}), we ran \texttt{run\_tasks.sh} twice, once per plot task, with the partition's profiles and parameters.
With \texttt{-{}-include-deps}, each invocation pulled in the builds and benchmarks that its plot needs, and in the second, \texttt{-{}-skip-succeeded} reused the builds that the first had completed.
Figure~\ref{fig:campaign} shows how the campaign maps onto Slurm jobs.

\subsection{Partitions, Parameters, and Environment}
\label{sec:eval-sites}
The campaign ran on three Slurm partitions of two supercomputers (Table~\ref{tab:sites}), with 455 benchmark runs in total.
We set the problem size and the parallelism levels for each node type on the command line, and every run recorded them together with the loaded modules.
During development, an earlier version of the campaign also ran on a 16-core workstation with the local profile, and those runs are not part of the reported data.

\begin{table*}[t]
  \caption{Partitions, parameters, and software of the case study, taken from the run records and the saved result sets, with five repetitions per point everywhere.}
  \label{tab:sites}
  \tablestyle
  \begin{tabularx}{\textwidth}{>{\bfseries}llrYrlrl}
    \toprule
    & & & & \multicolumn{2}{l}{\textbf{Single node}} & \multicolumn{2}{l}{\textbf{Multiple nodes}} \\
    \cmidrule(r){5-6} \cmidrule{7-8}
    Partition & \textbf{CPU (2 sockets per node)} & \textbf{Cores per node} & \textbf{Software} & \textbf{\boldmath$N$} & \textbf{Levels} & \textbf{\boldmath$N$} & \textbf{Nodes} \\
    \midrule
    \Otus & AMD EPYC 9655 (Turin) & 192 & GCC 14.2.0, OpenMPI 5.0.7, Julia 1.12.6 & 8192 & 1, 8, 16, 32, 64, 96, 128, 192 & 16384 & 1, 2, 4, 8, 16 \\
    \shaded \Kassel \texttt{pub23} & AMD EPYC 7443 (Milan) & 48 & GCC 14.2.0, OpenMPI 5.0.5, Julia 1.12.6 & 6144 & 1, 2, 4, 8, 16, 24, 48 & 12288 & 1, 2, 4, 8, 16 \\
    \Kassel \texttt{FB16} & Intel Xeon E5-2643 v4 & 12 & GCC 14.2.0, OpenMPI 5.0.5, Julia 1.12.6 & 4096 & 1, 2, 4, 6, 8, 12 & 8192 & 1, 2, 4, 8 \\
    \bottomrule
  \end{tabularx}
\end{table*}

The task files were identical on all three partitions: the results of each partition were saved with the git commit of the campaign and a patch of the uncommitted changes, the three commits do not differ in \texttt{tasks/}, and no patch touches \texttt{tasks/}.
The 37 task files in the artifact differ from their measured versions only in comments and in the default node list, which every multi-node invocation set on the command line.
What differed per partition were the profiles, the command-line parameters, and the Julia preferences generated on each login node (Section~\ref{sec:eval-lessons}).
Each partition has one profile for the C builds, the C/OpenMP runs, and the plots, a second for the Julia build and the single-node MPI jobs, and one per node count, plus a settings file shared by the node-count profiles.
Together these are 67 to 79 non-comment lines per partition, plus the sbatch template shared by all partitions (55 lines, 28 without comments).
The profiles set the partition, the resources, the modules, the paths of a per-user Julia installation, and the MPI launcher.

The five repetitions per point are statistical repetitions within one campaign, not repeats in the sense of Pearce et al.~\cite{pearce2025rrr}.
The runs on the two \Kassel partitions used different hardware and a different MPI build than those on \Otus, and the records show both.
Because we also adapted the problem size to each node type, they do not reproduce the measurements in that sense, and the results of different partitions are not directly comparable.
Our data contain no replicated run either, that is, a run on the same hardware with a different software stack.

\begin{figure*}[t]
  \centering
  \resizebox{\textwidth}{!}{\fontsize{8}{9.6}\selectfont\input{figs/eval-latex}}%
  \caption{
    Strong scaling of the three variants on one 192-core \Otus node ($N{=}8192$): execution time (left), speedup relative to one thread or rank (middle), and throughput (right).
    The bands in the time and throughput panels show the standard deviation over five repetitions.
    The figure was drawn from the aggregated data of the single-node plot task with the task's gnuplot script.
    Only its size and labels were adapted for the paper.
  }
  \label{fig:results}
\end{figure*}

\ifisc
\begin{figure*}[t]
  \centering
  \resizebox{\textwidth}{!}{\fontsize{8}{9.6}\selectfont\input{figs/eval-mn-latex}}%
  \caption{
    Strong scaling of the two MPI variants on 1 to 16 \Otus nodes with one rank per core ($N{=}16384$): execution time (left), speedup relative to one node (middle), and throughput (right).
    The bands in the time and throughput panels show the standard deviation over five repetitions.
    As in Figure~\ref{fig:results}, the figure was drawn from the aggregated data of the multi-node plot task with the task's gnuplot script.
    Only its size and labels were adapted for the paper.
  }
  \label{fig:results-mn}
\end{figure*}
\fi

\subsection{Results}
\label{sec:eval-results}
\ifisc
Figures~\ref{fig:results} and~\ref{fig:results-mn} show the single-node and multi-node results on \Otus.
\fi
On one node ($N{=}8192$), C/OpenMP throughput rises to 347~GFLOP/s at 96 threads, which fill one socket, and falls to 266~GFLOP/s at 192 threads.
The C/MPI and Julia/MPI variants track each other closely: both stay at about 28~GFLOP/s between 8 and 16 ranks and reach 262 and 259~GFLOP/s at 192 ranks.
At 192 cores, all three variants of this na\"ive kernel reach a parallel efficiency of 12 to 13\%.
Explaining the plateau and the drop beyond one socket is outside the scope of this paper.
Across nodes ($N{=}16384$), the two MPI variants reach speedups of 13.5 and 13.6 on 16 nodes (3,072 cores) relative to one node, an efficiency of 85\%.

\ifisc
The figures that the same task files produced on both \Kassel partitions are part of the artifact.
\fi
On these partitions, the multi-node points show relative standard deviations of up to 28\% (\texttt{pub23}) and 11\% (\texttt{FB16}), compared with less than 1\% on \Otus.
We leave the cause open.
The plot tasks also render a median-and-quartile version of every figure, which is less sensitive to such outliers~\cite{hoefler2015benchmarking}.

\subsection{Lessons from Porting}
\label{sec:eval-lessons}
Moving the campaign between supercomputers raised five practical problems.
We list them with the place in the campaign where each is handled.
\begin{itemize}
  \item \emph{Launcher.} Whether \texttt{srun} or \texttt{mpirun} starts the ranks as one MPI job depends on the supercomputer's MPI installation.
  The profile therefore chooses the launcher (\texttt{MPI\_LAUNCHER}, default \texttt{mpirun}), not the task.
  \item \emph{Placement.} Comparable measurements need the same placement in all variants, but OpenMP and MPI runtimes place threads and ranks by different defaults.
  The environment scripts therefore give all variants the same compact placement of one thread or rank per core, with defaults that a profile can override per supercomputer (\texttt{MPI\_\allowbreak LAUNCH\_\allowbreak OPTS}, \texttt{OMP\_PLACES}).
  \item \emph{Rank counts.} The partitions differ in cores per node, so for multi-node runs the \texttt{run\_benchmark} functions take the number of ranks from the allocation (\texttt{SLURM\_NTASKS}) and thus use all cores on every partition.
  Together with their \texttt{srun} branch, this is where the task files depend on Slurm, so a port to another workload manager would touch these functions as well.
  \item \emph{Login and compute nodes.} The manifest is built on the login node, whose core count can differ from that of the compute nodes, so the parallelism levels are given on the command line (\texttt{SCALING\_LEVELS}, \texttt{THREAD\_MAX}) instead of being derived from the login node.
  \item \emph{Offline compute nodes.} Compute nodes without network access cannot download Julia packages.
  The Julia project is therefore instantiated and bound to the supercomputer's MPI once on the login node, and the build run only checks the result.
  The run records flag the preferences file written there as an uncommitted change (Figure~\ref{fig:metadata}), so they document that the campaign depends on a generated file outside version control.
\end{itemize}
Most supercomputer-specific settings thus live in the profiles (\ref{p:profiles}) and in the command-line parameters, and the run records name the profile and capture the parameters and the loaded modules (\ref{p:records}).
The task files name no partition, walltime, or module, and their only Slurm dependencies are those listed above.
In short, a setting belongs in a profile if it describes the supercomputer, on the command line if it describes the experiment, and in a task file only if it holds on every supercomputer.

\section{Discussion and Limitations}
\label{sec:discussion}
By design, \Thinflow omits several features of the tools in Section~\ref{sec:related}.
It has no in-process data passing, no typed parameter schemas, and no task graphs that change shape during execution: the dependency graph is fixed in the Run Meta Phase, which suits benchmark sweeps but not workflows whose structure depends on intermediate results.

\paragraph{Relation to the established frameworks.}
For the case study, the comparison of Table~\ref{tab:concepts} can be made concrete.
The settings in our profiles, from the Slurm options to the MPI launcher, correspond to JUBE's platform files, ReFrame's site configuration, and Benchpark's system specification.
The problem sizes and parallelism levels that we set on the command line correspond to JUBE's parameter sets, ReFrame's test parameters, and Benchpark's experiment variables.
\Thinflow thus separates the same concerns as these frameworks.
It differs in that a task consists of plain shell files next to the code rather than of entries in an XML, YAML, or Python description, so its \texttt{run.sh} holds the commands that a job script would contain.
Conversely, the frameworks provide what \Thinflow leaves to its users, for example JUBE's analysers and result tables, ReFrame's comparison of measured performance with reference values, and Ramble's installation of the software stack and its inventory of hashes.
\Thinflow therefore serves authors whose campaigns start as job scripts, not groups whose supercomputers are already described in one of these frameworks.

\paragraph{Stages.}
All runs of a stage must finish before the next stage begins.
If one build takes seconds and another takes minutes, no benchmark starts before the slow build has finished.
In the case study, each build took about a second on \Otus.
With the Slurm profiles, a failed run also cancels all later stages, including runs that do not depend on it.
The dependencies themselves are declared per run, so a finer grouping would let each run start as soon as its own dependencies finish, at the cost of more jobs.

\paragraph{Shell.}
Task files look like the job scripts their authors already write, but sourcing them has pitfalls.
Under \texttt{set -e}, for example, a file whose last line is a false \texttt{[ \dots ] \&\& \dots} test returns a failure status and aborts the run, and the case study's task files warn about this in comments.
\texttt{-{}-dry-run} catches structural errors before anything is submitted: it resolves all tasks, runs, and dependencies and reports missing dependencies.
Errors in a task's own commands, however, appear only when that task runs.
Because the engine sources the campaign's shell files, a campaign is trusted code and should be read before it is run, like any downloaded build script.

\paragraph{State as files.}
Keeping state in run directories (\ref{p:tree}) is a design decision rather than a consequence of the implementation language: a directory can be inspected with standard tools, compared, copied in parts, and archived into a single file.
The price is that there is no query language beyond \texttt{-{}-status} and standard tools, and a rerun replaces earlier results unless they were committed.

\paragraph{Footprint.}
The engine has about 3,100 lines and is copied into each campaign (\ref{p:engine}), so updating the engine for a new campaign never changes an existing one.
When the campaign is kept in git, the engine is versioned with it, so the commit in each run record also identifies the engine.

\paragraph{Limitations.}
\Thinflow orchestrates and records runs but does not make software environments equivalent.
Running a campaign on another supercomputer can reproduce it in the sense of Pearce et al.\ only as far as the supercomputer offers comparable compilers and libraries or a container supplies them.
Proprietary software and configurations that the other supercomputer does not offer remain out of reach.
We tested \ref{p:profiles} only with Slurm, on three partitions.
Other workload managers need new profiles and changes to the case study's \texttt{run\_benchmark} functions.
The case study uses a single synthetic kernel, and the cost of planning very large campaigns in Bash is not yet measured.

\paragraph{Where it fits.}
For campaigns whose shape is known in advance, such as parameter sweeps and scaling studies, an engine that ships with the campaign keeps the artifact runnable without a service or framework installation.
\Thinflow can be combined with Spack~\cite{gamblin2015spack} through the environment scripts, with containers through \texttt{RUN\_\allowbreak CONTAINER}, with analysis tools such as Thicket~\cite{brink2023thicket} in plot tasks, and possibly with executors such as PSI/J~\cite{hategan2023psij} that could consume the manifest.

\section{Conclusion and Future Work}
\label{sec:conclusion}
We presented \Thinflow, a small Bash engine that organizes an HPC benchmarking campaign as a directory tree of tasks and runs.
In its campaign model (\ref{c:model}), settings are inherited along the tree, and a plain-text manifest states the planned execution before anything is submitted.
Its implementation (\ref{c:impl}) confines job submission to per-supercomputer profiles, so the engine is workload-manager agnostic, and it records the environment, hardware, parameters, git state, and outcome of every run.
Because the engine is copied into each campaign, the campaign directory can serve as the artifact.

In our case study (\ref{c:case}), 37 task files with 217 non-comment lines drove a three-variant strong-scaling study from builds to plots, with 455 benchmark runs on three Slurm partitions of two supercomputers, on up to 16 nodes.
Moving the campaign between supercomputers raised five practical problems, from the choice of the MPI launcher to compute nodes without network access.
Their fixes went into the profiles, the environment scripts, the command-line parameters, and a one-time setup on the login node, not into the benchmark code.
With these fixes, the same task files ran on all three partitions.
What differed per partition were the profiles (67 to 79 non-comment lines), the command-line parameters, and a generated Julia preferences file.
The run records name the profile, capture the parameters and the loaded modules, and flag the preferences file as an uncommitted change.
The case study thus suggests that a few conventions on top of plain shell files suffice to move such a campaign between supercomputers and to record what ran: a directory tree with inherited settings, a manifest, per-supercomputer profiles, and per-run provenance records.

Several directions remain open: profiles for other workload managers, a finer stage grouping that lets each run start as soon as its own dependencies finish, a measured comparison of the same campaign in JUBE or Benchpark, and a port of a complex existing artifact to \Thinflow.

\section*{Code and Data Availability}
\Thinflow\RepoFootnoteMark{} is open source.
The artifact of this paper\footnote{\url{https://gitlab.com/anonymous_authors/thinflow_artifact}} contains the campaign, the profiles, the run records and reference results of all three partitions, and the engine's regression suite.
\ifanon
\else
The environment section of the run records is reduced to the experiment's variables.
An archival snapshot with a persistent DOI will be published with the final version of this paper.
\fi

\section*{AI-Use Disclosure}
The authors used an AI assistant to help revise and polish the text.
The code, the experiments, and their results are the authors' work.
The authors take full responsibility for the content of this paper.

\ifanon
\else
    \section*{Acknowledgment}
    This research was partially funded by the Deutsche Forschungsgemeinschaft (DFG, German Research Foundation) under project numbers~\href{https://gepris.dfg.de/gepris/projekt/558599020?language=en}{558599020} and~\href{https://gepris.dfg.de/gepris/projekt/470527619?language=en}{470527619}.

    The authors gratefully acknowledge the computing time made available to them on the \href{https://www.hkhlr.de/de/universitaeten/universitaet-kassel}{University of Kassel Cluster} and on the high-performance computer \href{https://pc2.uni-paderborn.de/systems-and-services/otus}{Otus at the NHR Center Paderborn Center for Parallel Computing (PC2)}.
    This center is jointly supported by the Federal Ministry of Research, Technology and Space and the state governments participating in the National High-Performance Computing (NHR) joint funding program.
\fi

\printbibliography

\end{document}